\documentclass[reprint,notitlepage,superscriptaddress,preprintnumbers,nofootinbib,nobibnotes,amsmath,amssymb,aps,twocolumn]{revtex4-1}
\usepackage{graphicx,grffile}
\usepackage{dcolumn}
\usepackage{bm}
\usepackage{hyperref}
\usepackage{aas_macros,braket}
\makeatletter
\def\lst@lettertrue{\let\lst@ifletter\iffalse}
\makeatother

\def\widebreve#1{\mathop{\vbox{\m@th\ialign{##\crcr\noalign{\kern3\p@}%
      \brevefill\crcr\noalign{\kern3\p@\nointerlineskip}%
      $\hfil\displaystyle{#1}\hfil$\crcr}}}\limits}

\begin{document}

\title{Alcock-Paczynski effects on wide-angle galaxy statistics}
%---
\author{Maresuke Shiraishi}
\affiliation{Department of General Education, National Institute of Technology, Kagawa College, 355 Chokushi-cho, Takamatsu, Kagawa 761-8058, Japan}
%---
\author{Kazuyuki Akitsu}
\affiliation{Kavli Institute for the Physics and Mathematics of the Universe (WPI), UTIAS, The University of Tokyo, Chiba 277-8583, Japan}
%---
\author{Teppei Okumura} 
\affiliation{Academia Sinica Institute of Astronomy and Astrophysics (ASIAA), No. 1, Section 4, Roosevelt Road, Taipei 10617, Taiwan}
\affiliation{Kavli Institute for the Physics and Mathematics of the Universe (WPI), UTIAS, The University of Tokyo, Chiba 277-8583, Japan}
%---

\date{\today}

%\pacs{98.80.Cq}

\begin{abstract}
  
  The Alcock-Paczynski (AP) effect is a geometrical distortion in three-dimensional observed galaxy statistics. In anticipation of precision cosmology based on ongoing and upcoming all-sky galaxy surveys, we build an efficient method to compute the AP-distorted correlations of galaxy number density and peculiar velocity fields for any larger angular scale not relying on the conventionally used plane-parallel (PP) approximation. Here, instead of the usual Legendre polynomial basis, the correlation functions are decomposed using tripolar spherical harmonic basis; hence, characteristic angular dependence due to the wide-angle AP effect can be rigorously captured. By means of this, we demonstrate the computation of the AP-distorted correlations over the various scales. Comparing our results with the PP-limit ones, we confirm that the errors due to the PP approximation become more remarkable as the visual angle of separation between target galaxies, $\Theta$, enlarges, and especially for the density auto correlation, the error exceeds $10\%$ when $\Theta \gtrsim 30^\circ$. This highlights the importance of the analysis beyond the PP approximation.
  
\end{abstract}

\maketitle

%~~~~~~~~~~~~~~~~~~~~~~~~~~~~~~~~~~~~~~~~~~~~~~~~~~~~~~~~~~~~~~~~~~~~~~~~~~~~

%%%%%%%%%%%%%%%%%%%%%%%%%%%%%%%%%%%%%%%%%%%%%%%%%%%%%%%%%%%%%%%%%%%%%%%%%%%%%%
\section{Introduction}
%%%%%%%%%%%%%%%%%%%%%%%%%%%%%%%%%%%%%%%%%%%%%%%%%%%%%%%%%%%%%%%%%%%%%%%%%%%%%%

Clustering properties of galaxies are largely affected by the initial condition, the energy content of the Universe and the state of gravity. The two-point correlation function (2PCF) is a handy tool to quantify them in three-dimensional space, and the comparison between theoretical predictions and observational data has brought about precise understandings of the Universe so far.

For such studies, three-dimensional position of each galaxy is essential information, while observation information is limited to two-dimensional angular coordinate and a redshift. In practice, the former information is reconstructed from the latter one by assuming underlying geometry of the Universe. In this process, if the assumption differs from the true cosmology, the reconstructed three-dimensional position should also shift from the true one, inducing nontrivial distortions in the 2PCFs. Such geometrical distortions, called the Alcock-Paczynski (AP) effect \cite{Alcock:1979mp}, must be accurately evaluated and calibrated for precise data analysis, and can be served as a cosmological probe on their own right (e.g., \cite{Ryden:1995kg,Ballinger:1996cd,Matsubara:1996nf,Matsubara:1999du,Outram:2003ew,Blake:2011ep,Blazek:2013kfa,Li:2014ttl,Alam:2016hwk,Akitsu:2019avy}).

In the literature, the AP effect on the 2PCFs has been dealt with mostly under the plane-parallel (PP) approximation. In general, the 2PCF for a pair of target fields at positions ${\bf x}_1$ and ${\bf x}_2$ is characterized by the separation vector ${\bf x}_{12} \equiv {\bf x}_1 - {\bf x}_2$ and the two line-of-sight (LOS) directions $\hat{x}_1 \equiv {\bf x}_1 / |{\bf x}_1|$ and $\hat{x}_2 \equiv {\bf x}_2 / |{\bf x}_2|$. The PP approximation, i.e., the identification between the two LOS directions as $\hat{x}_1 = \hat{x}_2$, makes the computation of the 2PCFs much simpler, while the accuracy is worsened as the opening angle between $\hat{x}_1$ and $\hat{x}_2$, dubbed as $\Theta$, becomes larger. According to Refs.~\cite{Szalay:1997cc,Szapudi:2004gh,Yoo:2013zga,Castorina:2017inr,Taruya:2019xsf,Castorina:2019hyr,Shiraishi:2020vvj}, the PP approximation gives rise to more than $10\%$ errors on the original nondistorted 2PCFs for $\Theta \gtrsim 30^\circ$; thus, the same level of error is naturally expected also in the AP-distorted ones. Such a misestimation may become more crucial in the cosmological analysis based on upcoming all-sky surveys such as SPHEREx~\cite{Dore:2014cca}.

In light of these situations, this paper examines the AP effect not relying on the PP approximation. As for the number density field, there are already few previous theoretical works (e.g., \cite{Matsubara:1996nf,Matsubara:1999du,Bonvin:2013ogt}), and in observation the effect has been analyzed without the PP approximation via anisotropy of baryon acoustic oscillations (BAOs) \cite{Okumura:2008}. In this paper, we upgrade the formalism newly including the peculiar velocity statistics. The AP distortion gives rise to nontrivial angular dependence characterized by higher-order Legendre multipoles in the 2PCFs. To deal with this, we, for the first time, introduce the tripolar spherical harmonic (TripoSH) decomposition, which is a successful technique for computing the 2PCFs without the PP approximation and works well for various shapes of density, velocity and ellipticity statistics~\cite{Szalay:1997cc,Szapudi:2004gh,Papai:2008bd,Bertacca:2012tp,Yoo:2013zga,Raccanelli:2013dza,Shiraishi:2020nnw,Shiraishi:2020pea,Shiraishi:2020vvj}. In this paper, we show that the special angular structure due to the wide-angle AP effect can even be fully captured, and accordingly, the AP-distorted 2PCFs become rigorously computable.

After building the formalism, we demonstrate the computation of the AP-distorted density-density, density-velocity and velocity-velocity correlations under the specific setup, and see how their sizes and shapes change from the original non-distorted ones. In comparison with the PP-limit results, we find that the PP approximation causes larger errors for larger $\Theta$ (e.g., $\gtrsim 10\%$ for $\Theta \gtrsim 30^\circ$ in the density-density correlation). These indicate that the analysis not relying on the PP approximation will be indispensable in near future and our formalism makes it feasible.

This paper is organized as follows. In the next section, we review the linear theory description for the 2PCFs of the density and velocity fields.
The formulation of the AP-distorted 2PCFs and the application of the TripoSH decomposition to them are presented in Sec.~\ref{sec:AP} and Sec.~\ref{sec:TripoSH}, respectively. After showing numerical results of the AP-distorted 2PCFs in Sec.~\ref{sec:results}, we give conclusions in Sec.~\ref{sec:conclude}.

%%%%%%%%%%%%%%%%%%%%%%%%%%%%%%%%%%%%%%%%%%%%%%%%%%%%%%%%%%%%%%%%%%%%%%%%%%%%%%
\section{Linear theory} \label{sec:linear}
%%%%%%%%%%%%%%%%%%%%%%%%%%%%%%%%%%%%%%%%%%%%%%%%%%%%%%%%%%%%%%%%%%%%%%%%%%%%%%

Throughout this paper, we analyze the 2PCFs of two spinless fields: number density fluctuation [$\delta({\bf x}) \equiv n({\bf x}) / \bar{n}(x) - 1$] and the LOS component of peculiar velocity [$u({\bf x}) \equiv {\bf v}({\bf x}) \cdot \hat{x}$] of galaxies. Since we are interested in the correlations at large scales, we can simply use the linear theory representations as summarized bellow.

In the linear theory, these fields are linearly connected to the matter fluctuation in real space $\delta_{\rm m}$. The explicit forms read \cite{Hamilton:1997zq,Burkey:2003rk,Yoo:2013zga}
\begin{align}
  \begin{split}
    \delta({\bf x}) &= \int \frac{d^3 k}{(2\pi)^3} e^{i {\bf k} \cdot {\bf x}}
  \left[ 
b  - 
i \frac{\alpha}{kx} (\hat{k} \cdot \hat{x}) f  
+ (\hat{k} \cdot \hat{x})^2 f  \right] \delta_{\rm m}({\bf k}) , \\
%---
u({\bf x}) &= \int \frac{d^3 k}{(2\pi)^3} e^{i {\bf k} \cdot {\bf x}} \,
i \frac{aH}{k} (\hat{k} \cdot \hat{x}) f \delta_{\rm m} ({\bf k}) ,
\end{split} \label{eq:delta_u}
\end{align}
where $b$ is the linear bias parameter for the number density field, $a$ is the scale factor, $H$ is the Hubble parameter, $\alpha \equiv d \ln \bar{n}(x) / d \ln x +  2$ is the selection function of the galaxy sample, and $f$ is a function of the growth rate $D$, reading $f \equiv d \ln D / d \ln a$. Besides, $\delta_{\rm m}$ is given by a triple product of $D(x)$, the time-independent transfer function $T(k)$ and the primordial curvature perturbation $\zeta({\bf k})$; namely, $\delta_{\rm m}({\bf k}, x) = D(x) T(k) \zeta({\bf k})$. These parameters have dependence on time, redshift or comoving distance, while we do not often state it clearly as an argument for notational convenience. Let us adapt this manner to all variables henceforth unless the parameter dependence is nontrivial. Note that Eq.~\eqref{eq:delta_u} holds in redshift space. In real space, the velocity field's form remains unchanged, while the density one reduces to $\delta = b \delta_{\rm m}$ because of the absence of the redshift-space distortion terms.

Because of good compatibility with the later computations of the 2PCFs, by utilizing the Legendre polynomials ${\cal L}_\ell(y)$, let us rewrite Eq.~\eqref{eq:delta_u} into the unified form, reading
\begin{align} 
  X({\bf x}) =
  \int \frac{d^3 k}{(2\pi)^3} e^{i {\bf k} \cdot {\bf x}} 
  \sum_\ell  c_\ell^{X} (k)
  {\cal L}_\ell(\hat{k} \cdot \hat{x})
  \delta_{\rm m}({\bf k})
  , \label{eq:X}
\end{align}
where $X = \{\delta, u \}$ and 
\begin{align}
  \begin{split}
    c_\ell^\delta(k) &= \left( b + \frac{1}{3} f \right) \delta_{\ell,0}^{\rm K}
    - i \frac{\alpha}{k x} f \, \delta_{\ell,1}^{\rm K}
    + \frac{2}{3} f \, \delta_{\ell, 2}^{\rm K} ,\\
    %---
    c_\ell^u(k) &= i \frac{aH}{k} f \, \delta_{\ell, 1}^{\rm K} .
  \end{split} \label{eq:X_coeff}
\end{align}
with $\delta_{a,b}^{\rm K}$ denoting the Kronecker delta.

Throughout this paper, we impose the statistical homogeneous and isotropic condition of the matter power spectrum, so that
\begin{align}
  \Braket{\delta_{\rm m}({\bf k}_1) \delta_{\rm m}({\bf k}_2)} = (2\pi)^3 \delta^{(3)}({\bf k}_1 + {\bf k}_2) P_{\rm m}(k_1) . \label{eq:Pm}
\end{align}
From this and Eq.~\eqref{eq:X}, the form of the 2PCF, $\xi^{X_1 X_2}({\bf x}_1, {\bf x}_2) \equiv \Braket{X_1({\bf x}_1) X_2({\bf x}_2)}$, is derived, reading
\begin{align}
  \xi^{X_1 X_2}({\bf x}_1, {\bf x}_2)
  = \int \frac{d^3 k}{(2\pi)^3} e^{i {\bf k} \cdot {\bf x}_{12}} P^{X_1 X_2}({\bf k}, \hat{x}_1, \hat{x}_2) , \label{eq:xi}
\end{align}
where ${\bf x}_{12} \equiv {\bf x}_1 - {\bf x}_2$ and 
\begin{align} 
  P^{X_1 X_2}({\bf k}, \hat{x}_1, \hat{x}_2)
  &\equiv  \sum_{\ell_1 \ell_2} (-1)^{\ell_2} c_{\ell_1}^{X_1}(k) c_{\ell_2}^{X_2} (k) \nonumber \\
 &\quad \times {\cal L}_{\ell_1}(\hat{k} \cdot \hat{x}_1) 
  {\cal L}_{\ell_2}(\hat{k} \cdot \hat{x}_2) P_{\rm m}(k) . \label{eq:P}
\end{align}
Note that $P^{X_1 X_2}({\bf k}, \hat{x}_1, \hat{x}_2)$ is not equal to the Fourier counterpart of $\xi^{X_1 X_2}({\bf x}_1, {\bf x}_2)$ since there still remains the position dependence.

In the following, we reveal how the original 2PCFs~\eqref{eq:xi} are distorted by the AP effect.

%%%%%%%%%%%%%%%%%%%%%%%%%%%%%%%%%%%%%%%%%%%%%%%%%%%%%%%%%%%%%%%%%%%%%%%%%%%%%%
\section{Alcock-Paczynski effect} \label{sec:AP}
%%%%%%%%%%%%%%%%%%%%%%%%%%%%%%%%%%%%%%%%%%%%%%%%%%%%%%%%%%%%%%%%%%%%%%%%%%%%%%

Let us assume that the parameter vector describing the background geometry assumed for the reconstruction of the three-dimensional position of each galaxy ($\widetilde{\bf \Omega}$) is not equal to the true one (${\bf \Omega}$).%
\footnote{Readers should not confuse this ${\bf \Omega}$ with a solid angle.}
This causes the misestimation of the comoving radial distance toward each galaxy [$x(z, {\bf \Omega}) \to \tilde{x} \equiv x(z, \widetilde{\bf \Omega})$] and also both size and direction of the separation vector between each galaxy pair; thus, the resultant 2PCFs differ from the true ones. On the other hand, the LOS direction of each galaxy ($\hat{x}$) is independent of ${\bf \Omega}$ (due to the isotropic assumption of the background Universe) and hence remains undistorted.

We now investigate such AP-distorted 2PCFs without the PP approximation. Then let us consider an advantage in current precision cosmology that the gap between the assumed cosmology and the true one is slight; namely, ${\bf \Delta\Omega} \equiv \widetilde{\bf \Omega} - {\bf \Omega}$ is within sub percent of ${\bf \Omega}$. This justifies the analysis with the linear-order Taylor expansion in terms of ${\bf \Delta\Omega}$ or relevant small quantities. Regarding the real-space density field, the equivalent analysis has been done in the literature \cite{Bonvin:2013ogt}. In this paper, it is extended to the redshift-space density field and also the velocity one.

The distorted comoving radial distance of each galaxy can be evaluated as
\begin{align}
  \tilde{x} &\simeq x(z, {\bf \Omega})
  + \frac{\partial x(z, {\bf \Omega})}{\partial {\bf \Omega}} \cdot {\bf \Delta \Omega}
  \nonumber \\
 &\equiv x + \Delta x
  .
\end{align}
Since the gap of the comoving distance $\Delta x$ is also small enough compared to $x$, the AP-distorted field $X(\tilde{\bf x})$ can also be evaluated via the leading-order expansion around the true one $X({\bf x})$ in terms of ${\Delta x}$, so that
\begin{align}
 X( \tilde{\bf x})
  &\simeq X({\bf x}) + \frac{d X({\bf x})}{dx} \Delta x \nonumber \\ 
 & \equiv X({\bf x}) + \Delta X({\bf x}) . \label{eq:tilX}
\end{align}
Bearing in mind that $e^{i {\bf k} \cdot {\bf x}}$, $c_\ell^{X}$, and $\delta_{\rm m}$ in $X({\bf x})$ [Eq.~\eqref{eq:X}] depend on the comoving distance, the modulation term is computed as
\begin{align}
 & \Delta X({\bf x})
  = \int \frac{d^3 k}{(2\pi)^3} e^{i {\bf k} \cdot {\bf x}} 
  \sum_j  {\cal L}_j(\hat{k} \cdot \hat{x}) \delta_{\rm m}({\bf k}) \Delta x \nonumber \\
  &\qquad\times
  \left[ ik  (\hat{k} \cdot \hat{x}) c_j^{X} (k)   
    +  \frac{d \ln D}{d x} c_j^{X} (k)   
    + \frac{d c_j^{X} (k)}{dx}  
  \right].
\end{align}
Note that, here, following conventional analyses on the AP effect, we estimate a pure geometrical distortion due to the rescaling of the comoving distance; thus, a cosmological parameter shift in arguments of $\delta_{\rm m}$, $c_j^X$ and $D$ is not considered. Rewriting the angular-dependent part in the first term as 
\begin{align}
(\hat{k} \cdot \hat{x}) {\cal L}_{j}(\hat{k} \cdot \hat{x})
= \frac{4\pi}{3}
  \sum_{\ell} \frac{ h_{j 1 \ell}^2}{2j + 1} {\cal L}_{\ell}(\hat{k} \cdot \hat{x}) ,
\end{align}
we find the similar form to $X({\bf x})$ [Eq.~\eqref{eq:X}], reading
\begin{align}
  \Delta X({\bf x}) = \int \frac{d^3 k}{(2\pi)^3} e^{i {\bf k} \cdot {\bf x}}
  \sum_{\ell} \Delta c_\ell^{X} (k) {\cal L}_\ell (\hat{k} \cdot \hat{x})
 \delta_{\rm m}({\bf k}) , \label{eq:DX} 
\end{align}
where 
\begin{align}
  \Delta c_{\ell}^X(k) &\equiv
 \left[ \frac{4\pi}{3} i \sum_{j}  \frac{h_{j 1 \ell}^2}{2j+1} k c_{j}^{X} (k) \right. \nonumber \\ 
  & \left. \quad
    + \frac{d \ln D}{d x} c_{\ell}^{X} (k)
    + \frac{dc_{\ell}^{X} (k)}{dx}  \right]
 \Delta x
    , \label{eq:DX_coeff}
\end{align}
and
\begin{align}
h_{l_1 l_2 l_3} \equiv \sqrt{\frac{(2 l_1 + 1)(2 l_2 + 1)(2 l_3 + 1)}{4 \pi}} \left(\begin{matrix}
  l_1 & l_2 & l_3 \\
   0 & 0 & 0 
\end{matrix}\right),
\end{align}
with the function like the $2\times 3$ matrix denoting the Wigner $3j$ symbol. Since $h_{j 1 \ell}$ vanishes except for $|j - 1| \leq \ell \leq j + 1$ and $j + 1 + \ell = {\rm even}$ due to the selection rules of the Wigner $3j$ symbol, the number of nonvanishing multipoles is limited. The practical representations are as follows:
\begin{align}
  \begin{split}
  %--- delta
    \Delta c_\ell^{\delta} &= \left[
    \frac{1}{3} i k c_1^{\delta}  
    - aHf c_{0}^{\delta}
    +  \frac{dc_{0}^{\delta}}{dx}     
    \right] \Delta x \, \delta_{\ell, 0}^{\rm K} \\
  %@@@
  &\quad + \left[
    i k \left( c_0^{\delta} + \frac{2}{5} c_2^{\delta} \right)
    - aHf c_{1}^{\delta}
    +  \frac{dc_{1}^{\delta}}{dx}     
    \right] \Delta x \, \delta_{\ell, 1}^{\rm K} \\
  %@@@
  &\quad + \left[
    \frac{2}{3} i k c_1^{\delta}
    - aHf c_{2}^{\delta}
    +  \frac{dc_{2}^{\delta}}{dx}     
    \right] \Delta x \, \delta_{\ell, 2}^{\rm K} \\
    %@@@
  & \quad
  + \frac{3}{5} i k c_2^{\delta} \Delta x \, \delta_{\ell, 3}^{\rm K}
  , \\
  %--- u
  \Delta c_{\ell}^{u} &=
  \frac{1}{3} i k c_1^{u} \Delta x \, \delta_{\ell, 0}^{\rm K} \\
  %@@@
  &\quad + \left[
    - aHf c_{1}^{u}
    +  \frac{dc_{1}^{u}}{dx}     
    \right] \Delta x \, \delta_{\ell, 1}^{\rm K} \\
  %@@@
  &\quad + \frac{2}{3} i k c_1^{u} \Delta x \, \delta_{\ell, 2}^{\rm K} ,
  \end{split} \label{eq:DX_coeff_list}
\end{align}
where we have used a fact that $d \ln D(x) / d x = -aHf$. As seen in this, the first term in Eq.~\eqref{eq:DX_coeff} coming from $d e^{i {\bf k} \cdot {\bf x}} /dx$ produces nonvanishing Legendre multipoles not existing in $X({\bf x})$ [Eq.~\eqref{eq:X}], i.e., $\ell = 3$ in $\delta(\tilde{\bf x})$ and $\ell = 0, 2$ for $u(\tilde{\bf x})$. 

With Eq.~\eqref{eq:tilX}, the AP-distorted 2PCF is evaluated up to leading order in terms of $\Delta x_1$ and $\Delta x_2$, so that 
\begin{align}
  \Braket{X_1( \tilde{\bf x}_1) X_2( \tilde{\bf x}_2)}
&\simeq  \xi^{X_1 X_2}({\bf x}_1, {\bf x}_2)
+  \Delta \xi^{X_1 X_2}({\bf x}_1, {\bf x}_2) \nonumber \\
&\equiv \tilde{\xi}^{X_1 X_2}({\bf x}_1, {\bf x}_2) ,
\end{align}
where the original contribution $\xi^{X_1 X_2}({\bf x}_1, {\bf x}_2) = \Braket{ X_1({\bf x}_1) X_2({\bf x}_2)}$ is given by Eq.~\eqref{eq:xi}, and the modulation part reads
\begin{align}
  \Delta \xi^{X_1 X_2}({\bf x}_1, {\bf x}_2)
  &\equiv \Braket{ X_1({\bf x}_1) \Delta X_2({\bf x}_2)}
  + \Braket{ \Delta X_1({\bf x}_1) X_2({\bf x}_2)} \nonumber \\
  %@@@
  &= \int \frac{d^3 k}{(2\pi)^3} e^{i {\bf k} \cdot {\bf x}_{12}}
  \Delta P^{X_1 X_2}({\bf k}, \hat{x}_1, \hat{x}_2) , \label{eq:Dxi}
\end{align}
with
\begin{align}
 & \Delta P^{X_1 X_2}({\bf k}, \hat{x}_1, \hat{x}_2) \nonumber \\
  &\quad = \sum_{\ell_1 \ell_2} (-1)^{\ell_2}
  \left[c_{\ell_1}^{X_1}(k) \Delta c_{\ell_2}^{X_2} (k) + \Delta c_{\ell_1}^{X_1}(k) c_{\ell_2}^{X_2} (k) \right] \nonumber \\
  &\qquad \times {\cal L}_{\ell_1}(\hat{k} \cdot \hat{x}_1) 
  {\cal L}_{\ell_2}(\hat{k} \cdot \hat{x}_2) P_{\rm m}(k)
   . \label{eq:DP}
\end{align}
The PP-limit formula is derived further taking $\hat{x}_1 = \hat{x}_2$ in this equation, while we treat $\hat{x}_1$ and $\hat{x}_2$ as different vectors in order to estimate any wide-angle contribution.

The rescaling of the comoving distance also gives rise to a modification of volume, while this does not affect the 2PCF because it is dimensionless \cite{Ballinger:1996cd,Blazek:2013kfa,Bonvin:2013ogt}.

%%%%%%%%%%%%%%%%%%%%%%%%%%%%%%%%%%%%%%%%%%%%%%%%%%%%%%%%%%%%%%%%%%%%%%%%%%%%%%
\section{Tripolar spherical harmonic decomposition} \label{sec:TripoSH}
%%%%%%%%%%%%%%%%%%%%%%%%%%%%%%%%%%%%%%%%%%%%%%%%%%%%%%%%%%%%%%%%%%%%%%%%%%%%%%

The TripoSH decomposition is a fast and efficient technique to compute the 2PCFs not imposing the PP approximation~\cite{Szalay:1997cc,Szapudi:2004gh,Papai:2008bd,Bertacca:2012tp,Yoo:2013zga,Raccanelli:2013dza,Shiraishi:2020nnw,Shiraishi:2020pea,Shiraishi:2020vvj}. We here apply it to $\xi^{X_1 X_2}$~\eqref{eq:xi} and $\Delta \xi^{X_1 X_2}$~\eqref{eq:Dxi}.

Since the matter power spectrum takes the isotropic form as in Eq.~\eqref{eq:Pm}, the angular dependence in the 2PCF can completely be decomposed using the zero total angular momentum version of the TripoSH basis \cite{Varshalovich:1988ye,Shiraishi:2020nnw}:
\begin{align}
  {\cal X}_{\ell \ell_1\ell_2}(\hat{x}_{12},\hat{x}_1,\hat{x}_2)
  &\equiv \{ Y_{\ell}(\hat{x}_{12}) \otimes \{ Y_{\ell_1}(\hat{x}_1) \otimes Y_{\ell_2}(\hat{x}_2) \}_{\ell} \}_{00} \nonumber \\
  %@@@
  &= \sum_{m m_1 m_2} 
  (-1)^{\ell_1 + \ell_2 + \ell}
    \left( \begin{matrix}
      \ell_1 & \ell_2 & \ell \\
      m_1 & m_2 & m
    \end{matrix}  \right) \nonumber \\ 
    &\quad \times Y_{\ell m}(\hat{x}_{12}) Y_{\ell_1 m_1}(\hat{x}_1) Y_{\ell_2 m_2}(\hat{x}_2).  \label{eq:Xbasis_def}
\end{align}
The TripoSH decomposition of $\xi^{X_1 X_2}$ and $\Delta \xi^{X_1 X_2}$ are performed according to 
\begin{align}
  \begin{split}
  \xi^{X_1 X_2}({\bf x}_1, {\bf x}_2)
  &= \sum_{\ell\ell_1\ell_2} \Xi_{\ell\ell_1\ell_2}^{X_1 X_2}(x_{12}) 
     {\cal X}_{\ell\ell_1\ell_2}(\hat{x}_{12},\hat{x}_1,\hat{x}_2) , \\
     %---
  \Delta \xi^{X_1 X_2}({\bf x}_1, {\bf x}_2)
  &= \sum_{\ell\ell_1\ell_2} \Delta\Xi_{\ell\ell_1\ell_2}^{X_1 X_2}(x_{12}) 
         {\cal X}_{\ell\ell_1\ell_2}(\hat{x}_{12},\hat{x}_1,\hat{x}_2).
  \end{split} \label{eq:TripoSH_xi_def}
\end{align}
The decomposition coefficients $\Xi_{\ell\ell_1\ell_2}^{X_1 X_2}$ and $\Delta \Xi_{\ell\ell_1\ell_2}^{X_1 X_2}$ contain all the physical information. To compute them, let us predecompose $P^{X_1 X_2}$ and $\Delta P^{X_1 X_2}$ as
\begin{align}
  \begin{split}
    P^{X_1 X_2}({\bf k}, \hat{x}_1, \hat{x}_2)
    &= \sum_{\ell\ell_1\ell_2} \Pi_{\ell\ell_1\ell_2}^{X_1 X_2}(k) 
    {\cal X}_{\ell\ell_1\ell_2}(\hat{k},\hat{x}_1,\hat{x}_2) , \\
       %---
    \Delta P^{X_1 X_2}({\bf k}, \hat{x}_1, \hat{x}_2)
    &= \sum_{\ell\ell_1\ell_2} \Delta\Pi_{\ell\ell_1\ell_2}^{X_1 X_2}(k) 
    {\cal X}_{\ell\ell_1\ell_2}(\hat{k},\hat{x}_1,\hat{x}_2).
  \end{split}
\end{align}
Plugging this into Eqs.~\eqref{eq:xi} and \eqref{eq:Dxi} yields
\begin{align}
  \begin{split}
  \Xi_{\ell\ell_1\ell_2}^{X_1 X_2}(x_{12}) 
  &= i^{\ell} \int_0^\infty \frac{k^2 dk}{2\pi^2} j_{\ell}(k x_{12})
  \Pi_{\ell \ell_1 \ell_2}^{X_1 X_2}(k) , \\
  %---
  \Delta\Xi_{\ell\ell_1\ell_2}^{X_1 X_2}(x_{12}) 
  &= i^{\ell} \int_0^\infty \frac{k^2 dk}{2\pi^2} j_{\ell}(k x_{12})
  \Delta\Pi_{\ell \ell_1 \ell_2}^{X_1 X_2}(k) ,
  \end{split} \label{eq:hankel}
\end{align}
where $j_\ell(y)$ is the spherical Bessel function. The explicit forms of $\Pi_{\ell\ell_1\ell_2}^{X_1 X_2}$ and $\Delta\Pi_{\ell\ell_1\ell_2}^{X_1 X_2}$ are obtained computing 
\begin{align}
  \begin{split}
    \Pi_{\ell \ell_1 \ell_2}^{X_1 X_2}(k) 
    &=  \int d^2 \hat{k}  \int d^2 \hat{x}_1  \int d^2 \hat{x}_2 \,
    P^{X_1 X_2}({\bf k}, \hat{x}_1, \hat{x}_2) \\
    &\quad \times {\cal X}_{\ell \ell_1\ell_2}^{*}(\hat{k},\hat{x}_1,\hat{x}_2) , \\
    %---
    \Delta \Pi_{\ell \ell_1 \ell_2}^{X_1 X_2}(k) 
    &=  \int d^2 \hat{k}  \int d^2 \hat{x}_1  \int d^2 \hat{x}_2 \,
    \Delta P^{X_1 X_2}({\bf k}, \hat{x}_1, \hat{x}_2) \\
    &\quad \times {\cal X}_{\ell \ell_1\ell_2}^{*}(\hat{k},\hat{x}_1,\hat{x}_2) .
  \end{split}
\end{align}
The $\hat{x}_1$ and $\hat{x}_2$ integrals in both these equations have the same structure and are analytically reduced as
\begin{align}
 &  \int d^2 \hat{x}_1  \int d^2 \hat{x}_2 \,
   {\cal X}_{\ell \ell_1\ell_2 }^{ *}(\hat{k},\hat{x}_1,\hat{x}_2)
   {\cal L}_{j_1}(\hat{k} \cdot \hat{x}_1)
   {\cal L}_{j_2}(\hat{k} \cdot \hat{x}_2) \nonumber \\ 
   &\qquad = \frac{4\pi h_{\ell_1 \ell_2 \ell}}{(2\ell_1 + 1)(2\ell_2 + 1)}  
   \delta_{\ell_1, j_1}^{\rm K} \delta_{\ell_2, j_2}^{\rm K}.
  \label{eq:int_XLL}
\end{align}
We therefore obtain
\begin{align}
  \begin{split}
    \Pi_{\ell \ell_1 \ell_2}^{X_1 X_2}(k) 
    &= \frac{(4\pi)^2 (-1)^{\ell_2} h_{\ell \ell_1 \ell_2}}{(2\ell_1 + 1)(2\ell_2 + 1)} P_{\rm m}(k)  
    c_{\ell_1}^{X_1} (k) c_{\ell_2}^{X_2} (k) , \\
    %---
    \Delta \Pi_{\ell \ell_1 \ell_2}^{X_1 X_2}(k) 
    &= \frac{(4\pi)^2 (-1)^{\ell_2} h_{\ell \ell_1 \ell_2}}{(2\ell_1 + 1)(2\ell_2 + 1)} P_{\rm m}(k)  \\
 & \quad \times \left[ c_{\ell_1}^{X_1} (k) \Delta c_{\ell_2}^{X_2} (k)
      + \Delta c_{\ell_1}^{X_1} (k) c_{\ell_2}^{X_2} (k)
      \right] .
  \end{split}
  \label{eq:Pi}
\end{align}
Because of the selection rules of $h_{\ell \ell_1 \ell_2}$, i.e., $|\ell_1 - \ell_2|  \leq \ell \leq \ell_1 + \ell_2$ and $\ell + \ell_1 + \ell_2  = {\rm even}$, nonvanishing multipoles are restricted to 
\begin{align}
  \begin{cases}
    \Pi_{000}^{\delta \delta} , \ 
    \Pi_{011}^{\delta \delta} , \  
    \Pi_{022}^{\delta \delta} , \\
    %@@@
    \Pi_{101}^{\delta \delta} , \ 
    \Pi_{110}^{\delta \delta} , \ 
    \Pi_{112}^{\delta \delta} , \
    \Pi_{121}^{\delta \delta} , \\ 
    %@@@
    \Pi_{202}^{\delta \delta} , \ 
    \Pi_{211}^{\delta \delta} , \  
    \Pi_{220}^{\delta \delta} , \  
    \Pi_{222}^{\delta \delta} , \\
    %@@@
    \Pi_{312}^{\delta \delta} , \ 
    \Pi_{321}^{\delta \delta} , \\ 
    %@@@
    \Pi_{422}^{\delta \delta} ,
  \end{cases} 
    %---
  \begin{cases}
    \Pi_{011}^{\delta u} , \\
    %@@@
    \Pi_{101}^{\delta u} , \
    \Pi_{121}^{\delta u} , \\
    %@@@
    \Pi_{211}^{\delta u} , \\
    %@@@
    \Pi_{321}^{\delta u} ,
  \end{cases}
    %---
  \begin{cases} 
    \Pi_{011}^{u u} , \\
    %@@@
    \Pi_{211}^{u u} ,
  \end{cases}
  \label{eq:Pi_list}
\end{align}
and
\begin{align}
  \begin{split}
  %--- dd
 & \begin{cases}
     \Delta\Pi_{000}^{\delta \delta} , \ 
     \Delta\Pi_{011}^{\delta \delta} , \ 
     \Delta\Pi_{022}^{\delta \delta} ,
     \\
     \Delta\Pi_{101}^{\delta \delta} , \ 
     \Delta\Pi_{110}^{\delta \delta} , \ 
     \Delta\Pi_{112}^{\delta \delta} , \ 
     \Delta\Pi_{121}^{\delta \delta} , \ 
     \Delta\Pi_{123}^{\delta \delta} , \ 
     \Delta\Pi_{132}^{\delta \delta} , 
     \\
     \Delta\Pi_{202}^{\delta \delta} , \ 
     \Delta\Pi_{211}^{\delta \delta} , \ 
     \Delta\Pi_{213}^{\delta \delta} , \
     \Delta\Pi_{220}^{\delta \delta} , \ 
     \Delta\Pi_{222}^{\delta \delta} , \
     \Delta\Pi_{231}^{\delta \delta} , 
     \\
     \Delta\Pi_{303}^{\delta \delta} , \ 
     \Delta\Pi_{312}^{\delta \delta} , \ 
     \Delta\Pi_{321}^{\delta \delta} , \ 
     \Delta\Pi_{323}^{\delta \delta} , \
     \Delta\Pi_{330}^{\delta \delta} , \ 
     \Delta\Pi_{332}^{\delta \delta} , 
     \\
     \Delta\Pi_{413}^{\delta \delta} , \
     \Delta\Pi_{422}^{\delta \delta} , \
     \Delta\Pi_{431}^{\delta \delta} ,      
     \\
     \Delta\Pi_{523}^{\delta \delta} , \
     \Delta\Pi_{532}^{\delta \delta} , 
   \end{cases} \\
    %--- du
  & \begin{cases}
      \Delta\Pi_{000}^{\delta u} , \
      \Delta\Pi_{011}^{\delta u} , \
      \Delta\Pi_{022}^{\delta u} ,
     \\
     \Delta\Pi_{101}^{\delta u} , \
     \Delta\Pi_{110}^{\delta u} , \
     \Delta\Pi_{112}^{\delta u} , \
     \Delta\Pi_{121}^{\delta u} , 
     \\
     \Delta\Pi_{202}^{\delta u} , \
     \Delta\Pi_{211}^{\delta u} , \
     \Delta\Pi_{220}^{\delta u} , \
     \Delta\Pi_{222}^{\delta u} , \
     \Delta\Pi_{231}^{\delta u} ,
     \\
     \Delta\Pi_{312}^{\delta u} , \
     \Delta\Pi_{321}^{\delta u} ,
     \\
     \Delta\Pi_{422}^{\delta u} , \
     \Delta\Pi_{431}^{\delta u} ,
    \end{cases} \\
  %--- uu
  & \begin{cases}
      \Delta\Pi_{011}^{uu} ,
      \\
      \Delta\Pi_{101}^{uu} , \
      \Delta\Pi_{110}^{uu} , \
      \Delta\Pi_{112}^{uu} , \
      \Delta\Pi_{121}^{uu} ,  
      \\
      \Delta\Pi_{211}^{uu} , 
      \\
      \Delta\Pi_{312}^{uu} , \
      \Delta\Pi_{321}^{uu} . 
    \end{cases}
  \end{split} \label{DPi_list}
\end{align}
Here one can notice that $\Delta\Pi_{\ell \ell_1 \ell_2}^{X_1 X_2}$ becomes nonzero also at the multipoles where $\Pi_{\ell \ell_1 \ell_2}^{X_1 X_2}$ vanishes. This contributes to the generation of the slight difference in shape between $\xi^{X_1 X_2}$ and $\tilde{\xi}^{X_1 X_2}$ as seen in the next section.

Thanks to Eqs.~\eqref{eq:TripoSH_xi_def}, \eqref{eq:hankel} and \eqref{eq:Pi}, the precise computation of AP-distorted 2PCFs without the PP approximation becomes feasible. These are utilized for the numerical analysis presented in the next section.

%%%%%%%%%%%%%%%%%%%%%%%%%%%%%%%%%%%%%%%%%%%%%%%%%%%%%%%%%%%%%%%%%%%%%%%%%%%%%%
\section{Results} \label{sec:results}
%%%%%%%%%%%%%%%%%%%%%%%%%%%%%%%%%%%%%%%%%%%%%%%%%%%%%%%%%%%%%%%%%%%%%%%%%%%%%%

Now, we demonstrate the computation of the AP-distorted 2PCFs, $\tilde{\xi}^{X_1 X_2}({\bf x}_1, {\bf x}_2) = \xi^{X_1 X_2}({\bf x}_1, {\bf x}_2) + \Delta \xi^{X_1 X_2}({\bf x}_1, {\bf x}_2)$, beyond the PP limit. For this purpose, let us take the coordinate system adopted in Refs.~\cite{Szapudi:2004gh,Yoo:2013zga,Shiraishi:2020vvj} where the triangle formed with ${\bf x}_1$, ${\bf x}_2$ and ${\bf x}_{12}$ are confined to the $xy$ plane, and their directions are set to $\hat{x}_{12} = (1, 0, 0)$, $\hat{x}_1 = (\cos \phi_1 , \sin \phi_1 , 0)$ and $\hat{x}_2 = (\cos \phi_2 , \sin \phi_2 , 0)$ for $\phi_2 \geq \phi_1$. The opening angle between $\hat{x}_1$ and $\hat{x}_2$ is then given by $\Theta \equiv \phi_2 - \phi_1$. Note that this is equal to the opening angle between the wrongly assumed position vectors $\tilde{\bf x}_1$ and $\tilde{\bf x}_2$ since $\tilde{\bf x}_1 = \tilde{x}_1 \hat{x}_1$ and $\tilde{\bf x}_2 = \tilde{x}_2 \hat{x}_2$.

\begin{figure}[t]
  \begin{tabular}{c}
    \begin{minipage}{1.\hsize}
  \begin{center}
    \includegraphics[width = 1.\textwidth]{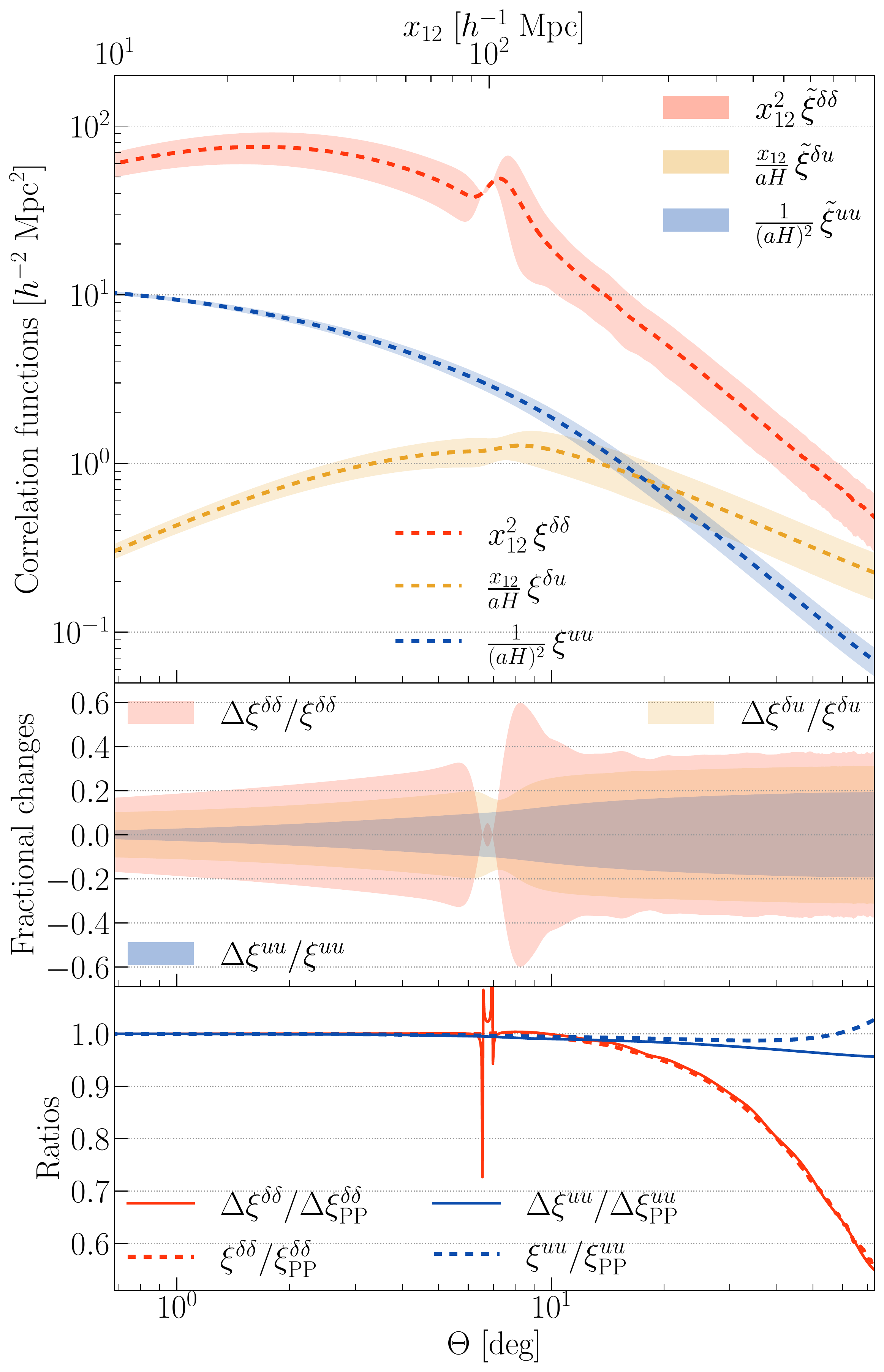}
  \end{center}
\end{minipage}
\end{tabular}
  \caption{Absolute values of the AP-distorted 2PCFs $\tilde{\xi}^{X_1 X_2}$ and the original ones $\xi^{X_1 X_2}$ (top), fractional changes of the AP-distorted 2PCFs from the original ones $\Delta \xi^{X_1 X_2} / \xi^{X_1 X_2}$ (middle) and ratios of our results to the conventional PP-limit ones $\Delta \xi^{X_1 X_2} / \Delta \xi_{\rm PP}^{X_1 X_2}$ and $\xi^{X_1 X_2} / \xi_{\rm PP}^{X_1 X_2}$ (bottom) as a function of $\Theta$ or $x_{12}$ at the equal-time or isosceles triangle condition ($z_1 = z_2 = 0.3$) and $b = 1$. Here, $\Delta \xi^{X_1 X_2}$ is estimated under a range of $-0.1 \leq \Delta x_1 / x_1 = \Delta x_2 / x_2 \leq 0.1$; thus, $\tilde{\xi}^{X_1 X_2}$ in the top panel and $\Delta \xi^{X_1 X_2} / \xi^{X_1 X_2}$ in the middle panel are described as the corresponding finite-width lines. Their upper (lower) boundaries correspond to the results for $\Delta x_1 / x_1 = \Delta x_2 / x_2 = -0.1 \, (0.1)$. Note that $\Delta \xi^{X_1 X_2} / \Delta \xi_{\rm PP}^{X_1 X_2}$ in the bottom panel is independent of $\Delta x_1 = \Delta x_2$ as both $\Delta \xi^{X_1 X_2}$ and $\Delta \xi_{\rm PP}^{X_1 X_2}$ are simply proportional to $\Delta x_1 = \Delta x_2$. At $x_{12} \simeq 100 h^{-1} \, {\rm Mpc}$, $\Delta \xi^{\delta \delta}$ and $\Delta \xi_{\rm PP}^{\delta \delta}$ vanish; thus, $\Delta \xi^{\delta \delta} / \Delta \xi_{\rm PP}^{\delta \delta}$ have some spiky features.}\label{fig:xi_per_Theta} 
\end{figure}

In Fig.~\ref{fig:xi_per_Theta}, we describe the results at the equal-time condition ($z_1 = z_2$), or equivalently, the isosceles triangle condition ($x_1 = x_2$ and $\Delta x_1 = \Delta x_2$). Then, $\phi_1$, $\phi_2$ and $x_{12}$ are solely specified by $\Theta$: $\phi_1 = (\pi - \Theta) / 2$, $\phi_2 = (\pi + \Theta) / 2$ and $x_{12} = x_1 \sqrt{2(1 - \cos\Theta)}$. Thus, $\xi^{X_1 X_2}({\bf x}_1, {\bf x}_2)$ and $\Delta \xi^{X_1 X_2}({\bf x}_1, {\bf x}_2)$ at a given redshift become a single-variate function of $\Theta$ or $x_{12}$. In this case, the wrongly estimated separation $\tilde{x}_{12}$ is related to the original true one $x_{12}$ via $\tilde{x}_{12} = x_{12} \, \tilde{x}_1 / x_1 = x_{12} \, \tilde{x}_2 / x_2$. We then consider within $10\%$ distortion of the comoving distance and hence move $\Delta x_1 / x_1$ and $\Delta x_2 / x_2$ from $-0.1$ to $0.1$. The values of the other relevant parameters adopted here are $z_1 = z_2 = 0.3$ and $b = 1$. The input cosmological parameters are fixed to be consistent with the latest {\it Planck} constraints \cite{Aghanim:2018eyx}.

The top and middle panels contain $\tilde{\xi}^{X_1 X_2}$, $\xi^{X_1 X_2}$ and $\Delta \xi^{X_1 X_2} / \xi^{X_1 X_2}$. When target galaxies look closer to us (i.e., $\Delta x_1 = \Delta x_2 < 0$), the physical size of any cosmological object should look more widened, and also the 2PCFs should look more amplified. These are visually apparent from the top panel, and especially from the slight shift of the BAO peak at $x_{12} \simeq 100 h^{-1} \, {\rm Mpc}$ to the upper right in the $\delta\delta$ case. Similarly, for the case of $\Delta x_1 = \Delta x_2 > 0$, the position and amplitude of the BAO peak change toward the opposite direction. The middle panel provides more quantitative information; namely, the ${\cal O}(1 - 10\%)$ enhancement/reduction depending mildly on $x_{12}$ or $\Theta$ is caused by a $10\%$ under/overestimation of the comoving distance. One can also find there a tendency that the velocity statistics are less distorted than the density ones.

The bottom panel plots the ratios to the PP-limit contributions, $\Delta \xi^{X_1 X_2} / \Delta \xi_{\rm PP}^{X_1 X_2}$ and $\xi^{X_1 X_2} / \xi_{\rm PP}^{X_1 X_2}$. Here, for the PP-limit condition $\hat{x}_{12} \cdot \hat{x}_1 = \hat{x}_{12} \cdot \hat{x}_2 = 0$, $\xi_{\rm PP}^{\delta u}$ and $\Delta \xi_{\rm PP}^{\delta u}$ vanish \cite{Burkey:2003rk,Howlett:2016urc,Shiraishi:2020pea} and therefore only the results for the $\delta \delta$ and $uu$ cases are displayed. From this, one can diagnose the reasonability of the conventional PP approximation. The errors on $\Delta \xi_{\rm PP}^{uu}$ and $\xi_{\rm PP}^{uu}$ remain within $10\%$ even at $\Theta \simeq 70^\circ$. In contrast, the errors on $\Delta \xi_{\rm PP}^{\delta \delta}$ and $\xi_{\rm PP}^{\delta \delta}$ grow more rapidly, and exceed $10\%$ for $\Theta \gtrsim 30^\circ$, highlighting the impact of beyond the PP approximation.

%%%%%%%%%%%%%%%%%%%%%%%%%%%%%%%%%%%%%%%%%%%%%%%%%%%%%%%%%%%%%%%%%%%%%%%%%%%%%%
\section{Conclusions}\label{sec:conclude}

In this paper, we have developed a new formalism for the AP-distorted 2PCFs of galaxy number density and peculiar velocity fields without the PP approximation. Since it is based on the TripoSH decomposition, the intricate angular dependence arising from the AP effect can fully be disentangled, achieving precise estimations of any larger angular scale contributions. Via numerical demonstration, we have revealed how the 2PCFs are distorted by the choice of a wrong geometry. Comparing our new results with the conventional PP-limit ones, it has also been confirmed that the errors induced by the PP approximation grow as the opening angle between two LOS directions $\Theta$ increases, and especially for the density autocorrelation, the error exceeds $10\%$ for $\Theta \gtrsim 30^\circ$. Therefore, the analysis not relying on the PP approximation is indispensable for precision cosmology based on ongoing and forthcoming wide-angle galaxy surveys, and our new formalism should be useful for it. Analyzing practical effects of the wide-angle AP distortion on the cosmological parameter search is beyond the scope of this paper, while it could be done by the application of the TripoSH-based Fisher matrix formalism~\cite{Shiraishi:2020nnw,Shiraishi:2020pea}.

Our results are based on the linear theory and the isotropic Universe assumption. On the other hand, general relativity, gravitational nonlinearity or cosmic isotropy breaking may produce the different shapes of the AP-distorted 2PCFs, and it will also be worth studying.

%%%%%%%%%%%%%%%%%%%%%%%%%%%%%%%%%%%%%%%%%%%%%%%%%%%%%%%%%%%%%%%%%%%%%%%%%%%%%

\acknowledgements

M.\,S. is supported by JSPS KAKENHI Grants No.~JP19K14718 and No.~JP20H05859. K.\,A. is supported by JSPS KAKENHI Grants No.~JP19J12254 and No.~JP19H00677. T.\,O. acknowledges support from the Ministry of Science and Technology of Taiwan under Grant No.~MOST 109-2112-M-001-027- and the Career Development Award, Academia Sinica (AS-CDA-108-M02) for the period of 2019 to 2023. M.\,S. and K.\,A. also acknowledge the Center for Computational Astrophysics, National Astronomical Observatory of Japan, for providing the computing resources of Cray XC50.

%%%%%%%%%%%%%%%%%%%%%%%%%%%%%%%%%%%%%%%%%%%%%%%%%%%%%%%%%%%%%%%%%%%%%%%%%%%%%%%%%%%%%%%%%%%%%%%%%%%%%%%%%%%%%%%%%%%%%%%%%%%%%%%%%%%%%%%%%%%%%%%%%%%%%%%%%%%%%%%%%%%%%%%%%%%%%%
% \appendix

%########################################
% Create the reference section using BibTeX
\bibliography{paper}
%\nocite{*}
%%%%%%%%%%%%%%%%%%%%%%%%%%%%%%%%%%%%%%%%%%

\end{document}